%% file: ms.tex
\documentclass[letterpaper,twocolumn,10pt]{article}
\usepackage{usenix,epsfig,endnotes}

\usepackage{url}
\usepackage{amsmath}
\usepackage{array}
\usepackage{amssymb}
\usepackage{amsthm}

\usepackage{threeparttable}
\usepackage{multirow}
\usepackage{hyperref}
\usepackage{breakurl}
 \hypersetup{
    colorlinks = true,
    linkcolor = red,
    anchorcolor = blue,
    citecolor = blue,
    filecolor = blue,
    urlcolor = blue
}

\usepackage{color}

\usepackage{fancyhdr}

\newtheorem{thm}{Theorem}

\begin{document}

\date{}

\title{\Large \bf Injected and Delivered: Fabricating Implicit Control over \\ Actuation Systems by Spoofing Inertial Sensors}


\author{
{Yazhou Tu$^*$ $\;\;\;\;\;\;$ Zhiqiang Lin$^\dag$$\;\;\;\;\;\;$  Insup Lee$^\ddag$$\;\;\;\;\;\;$   Xiali Hei$^*$} \\
$^*$University of Louisiana at Lafayette\\
$^\dag$The Ohio State University\\
$^\ddag$University of Pennsylvania
}


\maketitle

\renewcommand{\paragraph}[1]{\vspace{0.1in}\noindent{\bf{#1}.}}

\newcommand{\smallsection}[1]{\vspace{0.1in}\noindent{\emph{#1}.}}

\def\barr{\begin{tabular}{@{}l}}
\def\earr{\end{tabular}}

\def\barrc{\begin{tabular}{@{}c}}


\pagestyle{fancy}
\fancyhf{} 
\renewcommand{\headrulewidth}{0pt}
\fancyfoot[C]{This paper will appear in the proceedings of the USENIX Security Symposium, 2018.}
\fancyfoot[R]{\thepage}


\subsection*{Abstract}

Inertial sensors provide crucial feedback for control systems to determine motional status and make timely, automated decisions. Prior efforts tried to control the output of inertial sensors with acoustic signals. However, their approaches did not consider sample rate drifts in analog-to-digital converters as well as many other realistic factors. As a result, few attacks demonstrated effective control over inertial sensors embedded in real systems.

This work studies the out-of-band signal injection methods to deliver adversarial control to embedded MEMS inertial sensors and evaluates consequent vulnerabilities exposed in control systems relying on them. Acoustic signals injected into inertial sensors are out-of-band analog signals. Consequently, slight sample rate drifts could be amplified and cause deviations in the frequency of digital signals. Such deviations result in fluctuating sensor output; nevertheless, we characterize two methods to control the output: \emph{digital amplitude adjusting} and \emph{phase pacing}. Based on our analysis, we devise non-invasive attacks to manipulate the sensor output as well as the derived inertial information to deceive control systems. We test 25 devices equipped with MEMS inertial sensors and find that 17 of them could be implicitly controlled by our attacks. Furthermore, we investigate the generalizability of our methods and show the possibility to manipulate the digital output through signals with relatively low frequencies in the sensing channel.

\input{body-cps}

{\footnotesize \bibliographystyle{acm}
\bibliography{./ms}}

\end{document}

%% file: body-cps.tex
\section{Introduction} \label{section_introduction}

Sensing and actuation systems are entrusted with increasing intelligence to perceive the environment and react to it. Inertial sensors consisting of gyroscopes and accelerometers measure angular velocities and linear accelerations, which directly depict movements and orientations of a device.
Therefore, systems equipped with inertial sensors are able to determine motional status and make actuation decisions in a timely, automated manner.
While inertial sensing allows a control system to actuate in response to environmental changes promptly, errors of inertial measurements could result in instantaneous actuations as well.

Micro-electro-mechanical systems (MEMS) gyroscopes are known to be susceptible to resonant acoustic interferences \cite{castro2007influence, dean2011characterization,dean2007degradation,yunker2011underwater}.
Son et al. showed that a drone could be caused to crash by disturbing the gyroscope with intentional resonant sound \cite{son2015rocking}.
Furthermore, Trippel et al. investigated the data integrity issue of MEMS accelerometers under acoustic attacks \cite{trippel2017walnut}. While they gained adversarial control over exposed accelerometers, few attacks demonstrated effective control over embedded sensors. Thus, it remains unrevealed that to what extent attackers could exploit embedded inertial sensors and possibly control the systems relying on them.

To achieve adversarial control over inertial sensors embedded in real systems, we need to consider several realistic factors:
(a) \emph{Attack setting}. Biasing attacks in \cite{trippel2017walnut} were conducted on exposed sensors connected to an Arduino board, making the sampling process and real-time sensor data accessible to attackers. In contrast, our work studies non-invasive attacks, implying that attackers cannot physically alter the system and can only infer necessary information about the sensor from observable phenomena.
(b) \emph{Sample rate}. The exact sample rate of embedded sensors could be difficult to access, and we find that slight drifts in the sample rate may cause troubles to attackers.
(c) \emph{Actuating direction}. While Trippel et al. \cite{trippel2017walnut} manipulated a smartphone controlled RC car by inducing sensor outputs in only one direction, most systems rely on inertial measurements in both directions for control purposes. In this work, we develop generalizable methods that could manipulate inertial measurements of embedded sensors and trigger actuations of different kinds of control systems in both directions.

Acoustic signals injected at resonant frequencies of inertial sensors are usually out-of-band signals, which will be sampled by the analog-to-digital converter (ADC) with an insufficient sample rate. We characterize this kind of attacks as \emph{out-of-band signal injections}, presenting several important features: (1) \emph{Amplification of sample rate drifts}. We find that tiny drifts in the sample rate of an ADC could be amplified and cause more significant deviations in the frequency of the digital signal. Consequently, it could be difficult to induce and maintain a DC (Direct Current, 0 Hz) sensor output as in prior work \cite{trippel2017walnut}.
The resulting digital signal serves as noises due to its oscillating nature; nevertheless, we perceive following properties to control it.
(2) \emph{Adjustable digital amplitudes}. Distortions caused by undersampling allow amplitudes of different digital samples within one cycle of oscillation adjustable. (3) \emph{Phase pacing}. We find that a phase offset could be induced in the digital signal by switching the frequency of out-of-band analog signals.

Based on our analysis, we develop non-invasive attacks to manipulate the output of embedded inertial sensors as well as the derived information to deceive different kinds of control systems. We evaluate our attacks on 25 devices equipped with various models of inertial sensors from different vendors. Our experimental results show that 23 devices could be affected by acoustic signals and 17 of them are susceptible to implicit control.
Our attack demonstrations include maliciously actuating the motor of self-balancing human transporters, manipulating a user's view in virtual reality (VR) systems, spoofing a navigation system (Google Maps),
etc. We have uploaded the demos of our proof-of-concept attacks at \url{https://www.youtube.com/channel/UCGMX3ZbElV7BZYIX7RtF5tg}.

In summary, we list our contributions as follows:

\begin{itemize}
\vspace{-1.26mm}
\item We devise two sets of novel spoofing attacks (\emph{Side-Swing} and \emph{Switching} attacks) against embedded MEMS inertial sensors to manipulate sensor outputs and the derived inertial information. The attacks are non-invasive and could deliver implicit control to different kinds of real systems relying on inertial sensors.
\vspace{-1.96mm}
\item We evaluate our attacks on 25 devices and find that 23 of them can be affected by acoustic signals, presenting different control levels. Our proof-of-concept attacks demonstrate adversarial control over self-balancing, aiming and stabilizing, motion tracking and controlling, navigation systems, etc.
\vspace{-1.2mm}
\item We propose the out-of-band signal injection model and methods to manipulate the oscillating digitized signal when an analog signal is sampled with an insufficient sample rate. We investigate the generalizability of our methods with a case study showing that attackers could manipulate the oscillating digitized signal by sending signals with relatively low frequencies through a universal sensing channel.
\end{itemize}


\section{Inertial Sensors in Control Systems} \label{section_background}



MEMS inertial sensors use mechanical structures to detect inertial stimuli and generate electrical signals to depict it.
MEMS accelerometers detect linear accelerations with a mass-spring structure.
While MEMS gyroscopes use a similar structure to sense Coriolis accelerations $a_{Cor}$, an extra vibrating structure is used to drive the sensing mass with a velocity $v$, which is orthogonal to the sensing direction. The angular velocity $\omega$ causing the Coriolis acceleration can be derived by: $a_{Cor} = -2\omega \times v$.


\begin{figure}
\centering
  \includegraphics[scale=0.78]{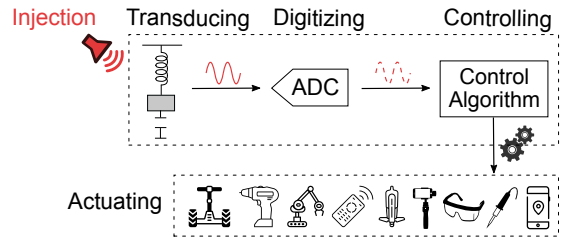}
  \vspace{-2mm}
  \caption{
An illustration of acoustic injections on inertial sensors embedded in control systems.
Injections of analog signals occur in the transducer. The signal will be digitized by the ADC before reaching the control system.
}\label{structure}
 \vspace{-1mm}
\end{figure}

\vspace{-1.2mm}

\paragraph{Acoustic Injection} Although MEMS technology has significantly reduced the size, cost and power consumption of inertial sensors, the miniaturized mechanical structure could suffer from resonant acoustic interferences. Acoustic signals at frequencies close to the natural frequency of the mechanical structure could force the sensing mass into resonance. Displacements of the sensing mass are usually measured by capacitive electrodes and would induce electrical signals.
The signal will then be digitized by the ADC and could possibly influence the control system, as shown in Figure \ref{structure}.

Under resonance, the sensing mass is forced into vibrations at the same frequency as the external sinusoidal driving force (sound pressure waves).
Therefore, the mass-spring structure of inertial sensors could serve as a receiving system for resonant acoustic signals and allow attackers to inject analog signals at specific frequencies.
However, the ability of attackers towards adversarial control is still restricted in two aspects: (1) Attackers cannot inject arbitrary forms of analog signals. Since the injected analog signal is caused by mechanical resonance of the sensing mass, it would be a sinusoidal signal and always presents an oscillating pattern.
(2) The digital signal cannot be controlled directly. Attackers could only induce specific digital signals by controlling the analog signal. This process is difficult to control especially in an embedded environment with limited information.

\vspace{-1.2mm}

\paragraph{Control System} MEMS inertial sensors provide crucial feedback for control systems to make autonomous decisions. Applications of MEMS gyros and accelerometers are very broad. Examples of these applications include human transporters, kinetic devices, robots, pointing systems for antennas, navigation of autonomous (robotic) vehicles, platform stabilization of heavy machinery, yaw rate control of wind-power plants, industrial automation units, and guidance of low-end tactical applications \cite{nasiri2009critical,antonello2011mems,passaro2017gyroscope,tian2016application}.
Because of their ubiquitousness and criticality in control systems, it is important to examine MEMS inertial sensors' reliability and evaluate the resilience of control systems under sensor spoofing attacks.





This work evaluates non-invasive spoofing attacks against embedded MEMS inertial sensors on a wide range of control systems in consumer applications. The systems we investigate can be broadly divided into two categories:
(1) \emph{Closed-loop control systems}.
The system continuously compares its current status with a goal status and tries to diminish the difference between them through actuations.
(2) \emph{Open-loop control systems}. The system simply follows inertial sensing information to make actuation decisions.
Different instances of these systems will be evaluated in Section \ref{section_manipulate}.

\vspace{-2mm}
\section{Threat Model} \label{section_threatmodel}
\vspace{-1mm}


The objective of attackers is to spoof embedded inertial sensors and deliver adversarial control to the system. To achieve this, attackers need to induce specific digital signals to trigger actuations in the control system.






\vspace{-1mm}

\paragraph{Non-invasiveness}
The spoofing attack against embedded inertial sensors is non-invasive and can be implemented without physical contact to the target device. Attackers cannot physically alter the hardware, neither can they directly access or modify the sampling process as well as the sensor output.
However, we assume that attackers can analyze the behavior of an identical device under acoustic effects before a real attack.

\vspace{-1mm}

\paragraph{Audibleness} The resonant frequencies of MEMS accelerometers are usually within the range of human hearing. However, the resonant frequencies of MEMS gyros are often in the ultrasound band (above 20 kHz). Therefore, acoustic signals used to attack gyros are inaudible. While resonant frequencies of gyros in several devices we test are between 19 to 20 kHz, they are still above the audible range of most adults \cite{takeda1992age}.




\vspace{-1mm}

\paragraph{Sound Source} Attackers can use consumer-grade speakers or transducers, directivity horns, and amplifiers to generate sound waves.
The signal source can be a function generator, an Arduino board, or mini signal generator boards \cite{mini_generator2, mini_generator}.
We assume that the possible attack distance is several meters;
attackers have sufficient resources, i.e., techniques or fund, to optimize the power, efficiency, directivity and emitting area of the sound source. More capable attackers could use professional acoustic devices or highly customized acoustic amplification techniques to further improve the range as well as the effect of the attack.

\vspace{-0.9mm}
\section{Modeling and Analysis} \label{section_theoretical}


In acoustic attacks, malicious analog signals injected into the transducer will be processed and digitized before reaching the control unit.
Therefore, the effect of attacks depends on the attacker's ability to influence the digitized signal.
In this section, we analyze the digitization process of out-of-band analog signals and propose general methods to control the oscillating digitized signal.







\subsection{Digitization of Out-of-band Signals } \label{subsection_digitizing}


Since the sensing mass under resonance is oscillating at the same frequency as sound waves, the resulting analog signal can be described by,

  \vspace{-2mm}
\begin{equation}\label{eq4_1}
\begin{array}{l}
V(t) = A\cdot sin(2\pi Ft+\phi_0)
\end{array}
  \vspace{-1mm}
\end{equation}

where $F$ is the frequency of resonant sound waves and the amplitude $A=A_0 k_a k_s$. $A_0$ is the amplitude of sound waves. The coefficients $k_a$ and $k_s$ represent the attenuation of acoustic energy during transmission and the sensitivity of the mechanical sensing structure respectively.
This analog signal will then be sampled by the ADC. Assuming $F_S$ is the sampling rate, and $t_0=0, t_1=\frac{1}{F_S},..., t_i=\frac{i}{F_S},...,$ are sampling times, the digitized signal will be,



\vspace{-3mm}
\begin{equation}\label{eq4_2}
\begin{small}
\begin{array}{l}
V[i] = A\cdot sin(2\pi F \frac{i}{F_S}+\phi_0) \ \ \ \ (i \in \{0, 1,2,3,...\})
\end{array}
\end{small}
  \vspace{-1mm}
\end{equation}

The frequency of analog signals injected through resonance is usually much higher than the sampling rate. For instance, the typical resonant frequency is several kHz for accelerometers and more than 19 kHz for gyros, while the sample rate is usually in tens or hundreds. According to the Nyquist theorem, when $ F > \frac{F_S}{2}$, there would be a problem of aliasing. We have,

  \vspace{-3mm}
\begin{equation}\label{eq4_3_pre_model}
\begin{array}{l}
F = n \cdot F_S + {\epsilon} \ \ \ \ (-\frac{1}{2}F_S < \epsilon \le \frac{1}{2}F_S, n \in \mathbb{Z}^+ )
\end{array}
  \vspace{-1mm}
\end{equation}

Substitute \eqref{eq4_3_pre_model} into \eqref{eq4_2}, we have:

  \vspace{-3mm}
\begin{equation}\label{eq4_4_basic_model_equation}
\begin{small}
\begin{array}{l}
V[i] = A\cdot sin(2\pi \epsilon \frac{i}{F_S}+\phi_0)  \ \ \ \ (i \in \{0, 1,2,3,...\})
\end{array}
\end{small}
  \vspace{-1mm}
\end{equation}


These equations describe the basic relationship between the out-of-band analog signal and the digitized signal: a sinusoidal analog signal with a frequency $F$ will be aliased to a digital signal with a frequency of $\epsilon$.



Our discussions in this section mainly focus on signals with frequencies close to the same integer multiple of sample rate. Therefore, we assume that $n$ in (\ref{eq4_3_pre_model}) stays the same when $\epsilon$, $F$ or $F_S$ slightly changes.

\begin{figure}
\centering
  \includegraphics[scale=0.77]{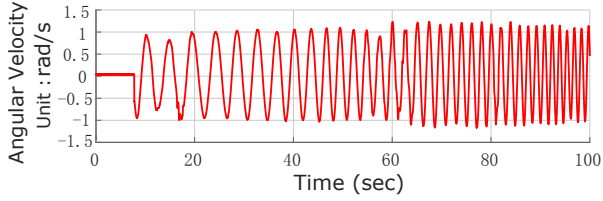}
  \vspace{-6.5mm}
  \caption{
The output of the gyroscope (X-axis) in a stationary iPhone 5S when we inject acoustic signals with a fixed frequency (19,471 Hz). Due to sample rate drifts, the frequency of the induced output is not a constant.
}\label{drifts1}
  \vspace{-3mm}
\end{figure}

\paragraph{Amplification Effect of Sample Rate Drifts} ADC is designed to sample the voltage of the analog signal at specific intervals. Theoretically, each interval should be exactly $\frac{1}{F_S}$. Therefore, given $F$, the value of $\epsilon$ should be determined (Equation \ref{eq4_3_pre_model}). However, due to drifts in $F_S$, when we inject acoustic signals at a fixed frequency into a smartphone's gyroscope, we find that the frequency of the digital output is deviating, as shown in Figure \ref{drifts1}. We formalize the following theorem to explain why slight sample rate drifts could result in observable deviations in the frequency of the digital signal.



\vspace{-1mm}

\begin{thm} When a signal with a frequency $F$ is sampled with an insufficient sample rate $F_S$ ($F_S < 2F$), a drift $\Delta_{F_S}$ in the sample rate will be amplified to a deviation of $-n \cdot \Delta_{F_S}$ in the frequency ($\epsilon$) of the sampled signal and $ n = \frac{F - \epsilon}{F_S } (-\frac{1}{2}F_S < \epsilon \le \frac{1}{2}F_S, n \in \mathbb{Z}^+ )$.
\end{thm}

\vspace{-3mm}

\begin{proof}
Let $\hat{\epsilon}$ be the frequency of the sampled signal after sample rate drifts. We have,

\vspace{-1.6mm}
\begin{equation}\label{dfas}
\begin{small}
\begin{array}{l}
F = nF_S + \epsilon \\
F = n(F_S+\Delta_{F_S}) + \hat{\epsilon}
\end{array}
\end{small}
 \vspace{-1mm}
\end{equation}

Therefore, the deviation in the frequency of the sampled signal is,

\vspace{-3.5mm}
\begin{equation}\label{dfas}
\begin{small}
\begin{array}{l}
\hat{\epsilon} - \epsilon = -n \cdot \Delta_{F_S}
\end{array}
\end{small}
  \vspace{-1mm}
\end{equation}


\end{proof}

  \vspace{-2mm}


For instance, the resonant frequency of gyros could range from 19 kHz to above 30 kHz. If $F=20,000$ Hz and $F_S=200$ Hz, a tiny drift of $0.01$ Hz in the sample rate would result in a deviation of $-1$ Hz in the frequency of the sampled signal. Due to the amplification effect of sample rate drifts, it is difficult to induce and maintain a DC output especially when the sensor is embedded.






\subsection{Digital Amplitude Adjusting}


The injected analog signal caused by mechanical resonance of the sensing mass is an oscillating sinusoidal signal. According to (\ref{eq4_4_basic_model_equation}), the resulting digital signal will also be oscillating (when $\epsilon \ne 0$).
However, an oscillating digital output induced in the sensor could be interpreted as noises or environmental interferences by the system, and its effect could be limited to disturbances or denial of service (DoS).
In this subsection, we investigate the possibility to modify the oscillating pattern of the digital signal by modulating the amplitude of analog signals.

\begin{figure}
\centering
  \includegraphics[scale=0.65]{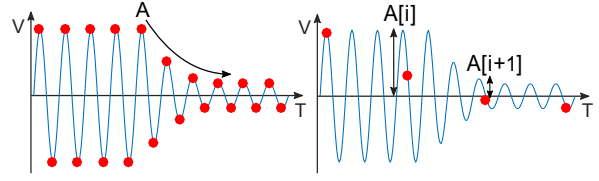}
  \vspace{-1.5mm}
  \caption{
When an oscillating analog signal is sampled correctly, the digital signal is oscillating (left). When an oscillating analog signal is undersampled, amplitudes of different digital samples could be adjusted to modify the shape of the digital signal (right).
}\label{discrete}
  \vspace{-3mm}
\end{figure}

An essential feature of out-of-band signal injections is that the induced analog signal will be undersampled, resulting in distortions of the signal. While aliasing is a well-known effect of signal distortions caused by undersampling, it mainly focuses on changes of the signal in the frequency domain, and how to utilize such distortions to intentionally modify the `shape' of an oscillating digitized signal has rarely been discussed.


Due to undersampling, the pattern of the analog signal may not be preserved in the digital signal.
As illustrated in Figure \ref{discrete}, when an amplitude modulated oscillating analog signal is sampled correctly, the digital signal has an amplitude that changes gradually and still presents an oscillating pattern. However, when an oscillating analog signal is undersampled, amplitudes of different digital samples within one cycle of oscillation ($T=\frac{1}{\epsilon}$) could be adjusted to modify the shape of the digital signal.
In fact, when $F > \frac{F_S}{2}$, the continuity in the amplitude of the oscillating analog signal kept in digitized samples begins to decrease.
As $\frac{2F}{F_S}$ grows, amplitudes of adjacent samples become less dependent on each other.
When $F$ is considerably larger than $\frac{F_S}{2}$, each digital amplitude can be adjusted independently.
We have,

\vspace{-5mm}
\begin{equation}\label{eq4_7_discrete}
\begin{small}
\begin{array}{l}
V[i] = A[i]\cdot sin(2\pi \epsilon \frac{i}{F_S}+\phi_0)   \ \ \ \ (i \in \{0, 1,2,3,...\})
\end{array}
\end{small}
\vspace{-1mm}
\end{equation}


where $A[0], A[1], A[2], ...$ could be adjusted by modulating the amplitude of the oscillating analog signal.
In this way, during out-of-band signal injections, a digital signal with specific waveforms (such as a one-sided waveform in Section \ref{subsection_sideswing}) instead of an oscillating signal could be fabricated.

\subsection{Phase Pacing}\label{Section_4_3_phasepacing}




In this subsection, we propose a novel approach to manipulate the phase of the oscillating digitized signal by changing the frequency of out-of-band analog signals.






Assuming the frequency of the analog signal changes from $F_1$ to $F_2$ at time $t_c$, and

\vspace{-5mm}
\begin{equation}\label{F1F2}
\begin{split}
F_1 & = n \cdot F_S + {\epsilon}_1 \quad \quad (-\frac{1}{2}F_S < \epsilon_1 \le \frac{1}{2}F_S, n \in \mathbb{Z}^+ )\\
\vspace{-3mm}
F_2 & = n \cdot F_S + {\epsilon}_2  \quad \quad (-\frac{1}{2}F_S < \epsilon_2 \le \frac{1}{2}F_S, n \in \mathbb{Z}^+ )
\end{split}
\vspace{-3mm}
\end{equation}

the analog signal will be:

\vspace{-5mm}
\begin{equation}\label{eq4_10_vts}
\begin{small}
\begin{array}{l}
 V(t)=

   \begin{cases}

    A\cdot sin(2\pi F_1 t+{\phi}_0) &\mbox{ $ 0 \le t \le t_c $ }\\

    A\cdot sin(2\pi F_2 ( t - t_c)+{\phi}_1) &\mbox{ $ t > t_c $}

   \end{cases}

\end{array}
\end{small}
\vspace{-1mm}
\end{equation}








where $\phi_0 $ is the initial phase of the analog signal, and $\phi_1$ is the phase of the analog signal when we change its frequency at $t_c$. We have:

\vspace{-2mm}
\begin{equation}\label{eq4_11}
\begin{array}{l}
{\phi}_1  = {2\pi F_1 t_c+{\phi}_0}
\end{array}
\vspace{-1mm}
\end{equation}

From (\ref{eq4_10_vts}) and (\ref{eq4_11}), we have,

\vspace{-3mm}
\begin{equation}
\begin{small}
\begin{array}{l}
 V(t)=

   \begin{cases}

   A\cdot sin(2\pi F_1 (t - t_c)+{\phi}_1)  &\mbox{ $ 0 \le t \le t_c $ }\\

    A\cdot sin(2\pi F_2 ( t - t_c)+{\phi}_1) &\mbox{ $ t > t_c $}

   \end{cases}
\end{array}
\end{small}
\vspace{-1mm}
\end{equation}









For simplicity, assuming $t_c = \frac{i_c}{F_s}$, the digitized signal will be,

\vspace{-5mm}
\begin{equation}\label{eq4_14}
\begin{array}{l}
V[i] = A\cdot sin(\Phi[i])  \ \ \ \ (i \in \{0, 1,2,3,...\})
\end{array}
\vspace{-1mm}
\end{equation}

where $\Phi[i]$ is the phase of the digital signal. We have,

\vspace{-5mm}
\begin{equation}\label{eq4_14_real}
\begin{small}
\begin{array}{l}

 \Phi[i]=

   \begin{cases}

     2\pi{\epsilon}_1 (\dfrac{i-i_c}{F_S}) + \phi_1 &\mbox{ $i \in \{0, 1, ...i_c\}$ }\\

    2\pi{\epsilon}_2 (\dfrac{i - i_c}{F_S}) + \phi_1 \hspace{-1mm} &\mbox{ $i \in \{i_c+1, i_c+2,...\}$}

   \end{cases}

\end{array}
\end{small}
\vspace{-1mm}
\end{equation}

Since $t_i = \frac{i}{F_S} $ is the sampling time, the derivative of the signal's phase will be

\vspace{-2mm}
\begin{equation}\label{eq4_15_derivative}
\begin{small}
\begin{array}{l}
 \Phi'[i]=

   \begin{cases}

     2\pi{\epsilon}_1  &\mbox{ $i \in \{0, 1, ...i_c\}$ }\\

    2\pi{\epsilon}_2  &\mbox{ $i \in \{i_c+1, i_c+2,...\}$}

   \end{cases}
\end{array}
\end{small}
\vspace{-1mm}
\end{equation}










Therefore, when the frequency of the analog signal changes at $t_c$, the phase of the signal is still $\phi_1$, but the derivative of the phase changes from
$2\pi\epsilon_1$ to $2\pi\epsilon_2$.
Especially, when

\begin{figure}
\centering
  \includegraphics[scale=0.8]{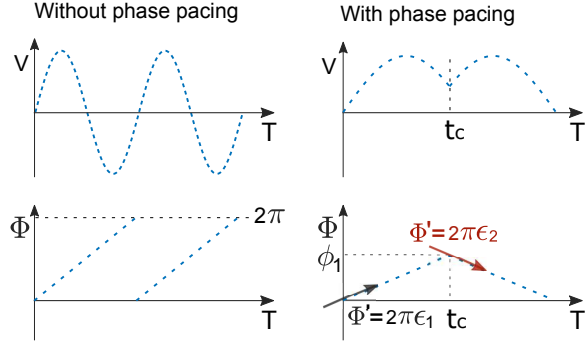}
  \vspace{-3mm}
  \caption{
Without phase pacing, the digital signal is oscillating (left).
With phase pacing at $t_c$, the moving direction of the digital signal is inverted due to the flipped sign of its phase derivative (right).
}\label{phase_pacing}
  \vspace{-1mm}
\end{figure}

\vspace{-4mm}
\begin{equation}\label{eq4_16_pacing_condition}
\begin{array}{l}

\epsilon_1 \cdot \epsilon_2 < 0,

\end{array}
\vspace{-1mm}
\end{equation}

the moving direction of the signal at $t_c$ will be inverted because of the flipped sign of the phase derivative, as illustrated in Figure \ref{phase_pacing}.







In fact, both parts of the digital signal can be represented in terms of positive frequencies. Assuming $\epsilon_1 > 0$, $\epsilon_2 < 0$, from (\ref{eq4_14}), (\ref{eq4_14_real}) and $sin(x)=sin(\pi-x)$, we have


\vspace{-5mm}
\begin{equation}
\begin{small}
\begin{array}{l}
 V[i]=

   \begin{cases}

   A\cdot sin(2\pi{\epsilon_1} (\dfrac{i-i_c}{F_S}) + \phi_1)  &\mbox{ $i \in \{0, 1, ...i_c\}$ }\\

    A\cdot sin(2\pi{(-\epsilon_2)} (\dfrac{i - i_c}{F_S}) + \pi - \phi_1) &\mbox{ $i \in \{i_c+1,...\}$}

   \end{cases}
\end{array}
\end{small}
\end{equation}

We can see clearly there is a phase change of $\pi-2\phi_1$ in the digital signal because of frequency switching at time $t_c$. We refer to the method that induces a phase offset in the digital signal by switching the frequency of out-of-band analog signals as \emph{Phase Pacing}.







\subsection{Out-of-band Signal Injection Model} \label{subsection_modeling}


In summary, during out-of-band signal injections, the digitized signal can be represented by,

\vspace{-5mm}
\begin{equation}\label{eq4_17_injection_model_a}
\begin{array}{l}
V[i] = A[i]\cdot sin(\Phi[i])   \ \ \ \ (i \in \{0, 1,2,3,...\})
\end{array}
\vspace{-1mm}
\end{equation}

Where,

\vspace{-5mm}
\begin{equation}\label{eq4_18_injection_model_p}
\begin{array}{l}
\Phi[i] = 2\pi \epsilon \frac{i}{F_S}+\phi_0   \ \ \ \ (i \in \{0, 1,2,3,...\})
\end{array}
\vspace{-1mm}
\end{equation}

The parameters that could be manipulated in this model are $A[i]$ and $\epsilon$. By adjusting $A[i]$, the value of each digitized sample $V[i]$ can be manipulated proportionally. In addition, $\epsilon$ can be altered by changing the frequency of the analog signal. Especially, when the sign of $\epsilon$ is flipped, the moving direction of the digital signal will be inverted because of the phase offset.




  \vspace{-0.5mm}
\section{Attack Methods} \label{section_attack}

Inertial sensors are often used by control systems to ascertain the state of motion. One critical property derived from inertial measurements is the heading angle. A different heading angle detected by the control system often triggers different automated decisions and actuations. Therefore, in this section, we investigate attack methods on embedded inertial sensors to manipulate sensor readings as well as the derived heading angle.

\subsection{Side-Swing Attack} \label{subsection_sideswing}

The basic idea of Side-Swing attacks is to proportionally amplify the induced output in the target direction and attenuate the output in the opposite direction.

\begin{figure}
\centering
  \includegraphics[scale=0.7]{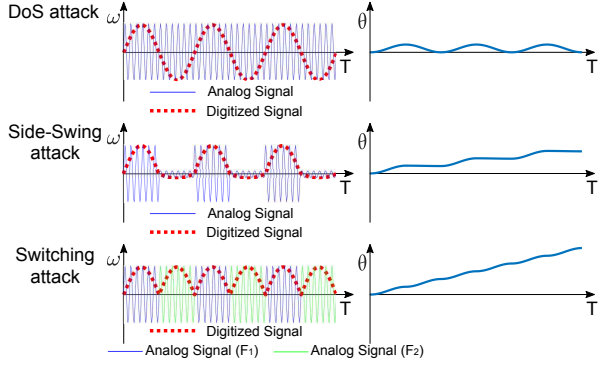}
  \vspace{-2mm}
  \caption{
For an oscillating signal, the accumulative heading degree ($\theta$) fluctuates and falls back to 0 after each cycle (top).
Under Side-Swing attacks, the derived heading degree grows but only in half of each period of the signal (middle).
The derived heading degree under Switching attacks keeps growing (bottom).
}\label{swing_principle_dos}
  \vspace{-2mm}
\end{figure}

In DoS attacks, the potential accumulative inertial information induced is often limited because an oscillating signal contributes to about the same amount of inertial measurements in both directions.
As illustrated in Figure \ref{swing_principle_dos}, when an oscillating sensor output is induced in a gyro, the heading angle $\theta$ accumulated in each cycle of oscillation is 0.

%

%
To address this problem, in Side-Swing attacks, the attacker can increase the amplitude when the digitized sample is in the target direction and decrease the amplitude otherwise.
Recall in (\ref{eq4_17_injection_model_a}), we have
$V[i] = A[i]\cdot sin(\Phi[i]) $. Assuming that the target direction is the positive direction, the attacker would increase $A[i]$ when $sin(\Phi[i]) > 0$, otherwise decrease $A[i]$ to 0 or a very small value.
In this way, the derived heading angle can be accumulated in the target direction. 



Assuming that the injected analog signals are modulated with a high amplitude $A_h$ and a low amplitude $A_l$ alternatively, the heading angle accumulated in each cycle of the signal will be,

\vspace{-4mm}
\begin{equation}\label{eq_degree}
\begin{small}
\begin{array}{l}
\theta = \int_0^{\frac{1}{2\epsilon}} A_h \cdot sin(2\pi \epsilon t ) + \int_{\frac{1}{2\epsilon}}^{\frac{1}{\epsilon}} A_l \cdot sin(2\pi \epsilon t ) = \frac{A_h-A_l}{\pi\epsilon} \\
\end{array}
\end{small}
\vspace{-0.5mm}
\end{equation}

The average angular speed during one cycle is:

\vspace{-3.0mm}
\begin{equation}\label{eq_degree}
\begin{array}{l}

\bar{\omega} =  \epsilon \theta = \frac{A_h-A_l}{\pi} \\
\end{array}
\vspace{-0.2mm}
\end{equation}

When $A_l=0$, the heading angle accumulated in one cycle would be $\frac{A_{h}}{\pi\epsilon}$, and the average angular velocity would be $ \frac{A_h}{\pi} $. Attackers can adjust these values by adopting different values of $A_h$. The principle of Side-Swing attacks is illustrated in Figure \ref{swing_principle_dos}.

\begin{figure}
\centering
  \includegraphics[scale=1.15]{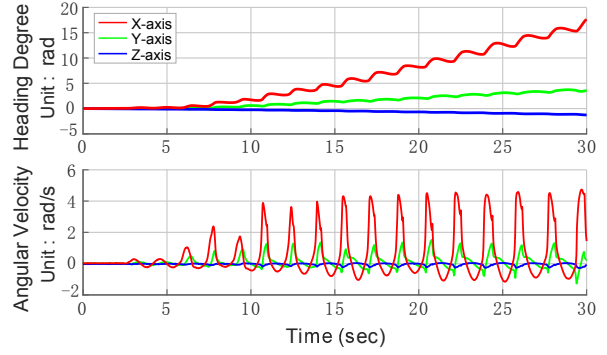}
  \vspace{-3.5mm}
  \caption{
Output of the gyroscope in an iPhone 5 and the derived heading angle under Side-Swing attacks in X-axis. The phone is 0.5 m away from a 50-Watt sound source. The sound frequency is 19,976 Hz.
}\label{swing_gyro_data}
  \vspace{-2mm}
\end{figure}

We conduct Side-Swing attacks on the gyroscope of an iPhone 5. As shown in Figure \ref{swing_gyro_data}, while the phone is stationary, the collected gyroscope data shows that it has rotated to the positive direction of X-axis for 17.6 rads (1008$^\circ$) in about 25 seconds. The peak angular speed $\omega_{max}$ is 4.73 rad/s and the average angular speed $\bar{\omega}$ is 0.70 rad/s. The ratio of $\bar{\omega}$ to $\omega_{max}$ is 0.15.



In summary, Side-Swing attacks induce the outputs mainly in the target direction and allow the derived heading angle to be manipulated.
In control systems, the moving direction and speed of actuators are often determined by the measured angular velocity and the derived heading angle.
Therefore, Side-Swing attacks could provide attackers a more direct way to manipulate the control system by modulating the amplitude of acoustic signals.
However, during Side-Swing attacks, the derived heading angle increases in only half of each period of the signal and stops growing when the signal is in the opposite direction. This may limit the maximum heading angle accumulated in a certain amount of time.

\subsection{Switching Attack} \label{subsection_switching}

The principle of Switching attacks is to control the induced output by manipulating the phase of the digital signal with repetitive phase pacing.


Recall (\ref{F1F2}) and (\ref{eq4_16_pacing_condition}) in Section \ref{Section_4_3_phasepacing}, when $\epsilon_1 \cdot \epsilon_2 < 0$ and the frequency of the analog signal changes from $F_1$ to $F_2$, the moving direction of the digital signal will be inverted. Similarly, if the frequency of the analog signal changes from $F_2$ to $F_1$, the condition of phase pacing ($\epsilon_2 \cdot \epsilon_1 < 0$) also holds. Therefore, in Switching attacks, the attacker uses two frequencies ($F_1$ and $F_2$) and switches the frequency of acoustic signals between them to induce phase pacing repeatedly. Different from Side-Swing attacks, the accumulated heading angle in Switching attacks keeps growing under the sustained influence of the induced angular speed in the target direction, as illustrated in Figure \ref{swing_principle_dos}.




Assuming the target direction is the positive direction and the attacker switches the frequency when the signal drops from the target direction to the opposite direction, the heading degree accumulated in one period would be:

\vspace{-3.6mm}
\begin{equation}\label{eq_degree2}
\begin{small}
\begin{array}{l}
\theta = \int_0^{\frac{1}{2\epsilon}} A \cdot sin(2\pi \epsilon t ) + \int_{\frac{1}{2\epsilon}}^{\frac{1}{\epsilon}} A \cdot sin(-2\pi \epsilon t + \pi) = \frac{2A}{\pi\epsilon} \\
\end{array}
\end{small}
\vspace{-0.9mm}
\end{equation}

where we assume $\epsilon_1 > 0$, $\epsilon_2 < 0$ and $|\epsilon_1| = |\epsilon_2| = \epsilon$ to simplify the discussion. The average angular speed in one period of the signal is

\vspace{-5mm}
\begin{equation}
\begin{array}{l}

\bar{\omega} =  \epsilon \theta = \frac{2A}{\pi} \\
\end{array}
\vspace{-1.2mm}
\end{equation}



\begin{figure}
\centering
  \includegraphics[scale=0.69]{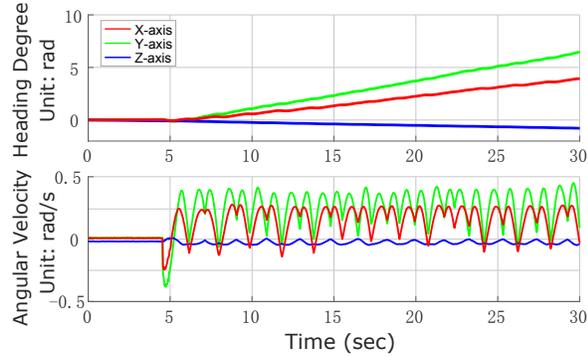}
  \vspace{-3.5mm}
  \caption{
Output of the gyroscope in an iPhone 7 and the derived heading angle under Switching attacks in Y-axis. The phone is 0.3 m away from a 50-Watt sound source. The sound frequencies are 27,378 and 27,379 Hz.
}\label{switching_gyro_data}
  \vspace{-8mm}
\end{figure}

The values of $\theta$ and $\bar{\omega}$ can be adjusted by adopting different amplitudes. In fact, the attacker can switch the frequency more frequently to keep the signal at a higher level and induce a larger heading angle. As shown in Figure \ref{switching_gyro_data}, we conduct Switching attacks on the gyroscope of an iPhone 7. While the phone is stationary, the collected gyroscope data shows that it has rotated to the positive direction of Y-axis for 6.5 rads (372.4$^\circ$) in about 25 seconds. The peak angular speed $\omega_{max}$ is 0.45 rad/s and the average angular speed $\bar{\omega}$ is 0.26 rad/s. The ratio of $\bar{\omega}$ to $\omega_{max}$ is 0.58, which is much larger than 0.15 in the previous experiment with Side-Swing attacks, implying that Switching attacks are more efficient than Side-Swing attacks and could be used to achieve a larger heading angle. However, acoustic frequencies used in Switching attacks should satisfy (\ref{F1F2}) and (\ref{eq4_16_pacing_condition}).
We can assume $F_2 = F_1 + step \ \  (F_1 < F_2)$, and the parameter $step$ can be selected by the attacker to control the length of the interval $[F_1, F_2]$ that bounds the integer multiple of $F_S$. In our settings, $step$ is set to $1$.


In summary, both Side-Swing and Switching attacks could induce spoofed sensor outputs in the target direction and manipulate the derived heading angle.
The target direction can be either positive or negative, determined by the attacker.
Theoretically, these methods are not limited to controlling oscillating digitized signals with a very small $|\epsilon|$.
However, in practice, the value of $|\epsilon|$ should be less than 0.5 or 1, depending on the reaction speed of an attacker.
With a very large $\epsilon$, the signal would oscillate rapidly and may allow not enough time to manually tune acoustic signals effectively.
Since the frequency ($\epsilon$) of the induced signal is closely related to the behavior of the device under attacks, we assume attackers could analyze the behavior of an identical device under acoustic effects to find suitable sound frequencies that could be used in the attack.








\vspace{-1mm}
\section{Evaluations} \label{section_manipulate}
\vspace{-0.5mm}






MEMS inertial sensors are widely used in consumer, industrial, and low-end tactical control systems \cite{nasiri2009critical,passaro2017gyroscope}. Depending on the application, the control algorithm and usage of inertial sensors might be different.
Therefore, a key question is: \emph{Can non-invasive spoofing attacks on embedded inertial sensors deliver adversarial control to various types or just one particular type of systems?} The answer to this question will give us a clearer understanding of the potential attack scope and facilitate the evaluation of vulnerabilities that might ubiquitously exist in control systems relying on MEMS inertial sensors.

We evaluate the non-invasive attacks on various types of real systems equipped with MEMS inertial sensors.
The results of our attack experiments are summarized in Table \ref{table_close_loop} and Table \ref{table_open_loop}.
Among the 25 tested devices, 17 devices are susceptible to implicit control. In remaining devices, 2 of them can be controlled very limitedly due to insufficient sound strength and 4 of them are vulnerable to DoS attacks. Only 2 devices are not affected by acoustic signals.
Our proof-of-concept attacks demonstrate implicit control over various systems including self-balancing, aiming and stabilizing, motion tracking and controlling, navigation systems, etc. 

In our experiments, we find that attacks on gyros induce more responsive actuations in the system and demonstrate more adversarial control than attacks on accelerometers. Possible reasons could be that gyros are usually more sensitive, and in most control systems with both gyros and accelerometers, the heading angle of the device is mainly derived from angular velocities measured by gyros, while accelerometers are often used as a gravity sensor and could slowly calibrate the derived orientation information.

\vspace{-1mm}
\subsection{Attack Overview}\label{subsection_overview}
\vspace{-0.5mm}


Without accessing the real-time inertial sensor data, it could be difficult for attackers to decide when to change the amplitude or frequency of acoustic signals so that malicious sensor data is induced in the target direction. However, we find that decisions made by control systems could give away certain information about the induced digital signal, and such information could be observed and leveraged to guide the attack.

\begin{figure}
\centering
  \includegraphics[scale=0.9]{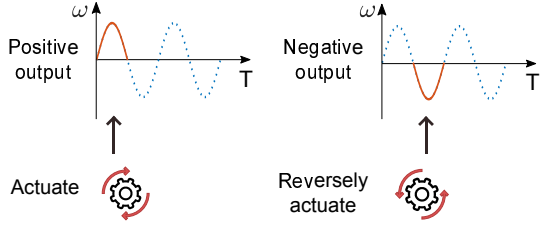}
  \vspace{-2.5mm}
  \caption{
An illustration of the reverse signal mapping method. Attackers could reversely infer the current direction and amount of the induced sensor output by observing the consequent actuations or accelerations.
}\label{reversemapping}
  \vspace{-4mm}
\end{figure}

During attacks, the induced sensor output could influence actuation decisions of the system instantaneously.
For instance, when positive sensor output is detected in the X-axis of the embedded gyro, a self-balancing human transporter would apply forward accelerations to the motor, while negative angular velocities would trigger accelerations to the opposite direction.
The amount of the induced acceleration is related to the amount of the spoofed angular velocity.
In turn, by observing consequent actuations or accelerations in the system, attackers could estimate the current direction and amount of the induced sensor output, as illustrated in Figure \ref{reversemapping}.
Another property that could be observed and estimated is the frequency ($|\epsilon|$) of the induced signal, which could be reversely mapped from the frequency of oscillating movements induced in actuation systems. Such oscillating movements could be periodic accelerations and decelerations of a motor, shaking or circling movements of visual information in VR/AR systems, etc.

The reversely inferring method could be used in following steps to guide the attack:


\emph{1) Profiling.} Before the attack, attackers could analyze the behavior of an identical device under acoustic effects to find the resonant frequency range and profile suitable attack frequencies of the embedded inertial sensor.


To find the resonant frequency range, attackers could generate single-tone sound and sweep a frequency range at an interval of 10 Hz.
Attackers apply the sound to a device that is stationary or in a well-balanced status, and there is no other input to control or interfere with the target system. The range of sound frequencies that noticeably affect the motion sensing unit and induce actuations in the device can be recorded as the resonant frequency range.
We notice that acoustic frequencies in the middle part of the range could affect the target device more significantly since they are closer to the natural frequency.

Attackers could then generate single-tone sound in the resonant frequency range and adjust the frequency with an interval of 1 Hz or smaller to find and profile attack frequencies.
Acoustic frequencies used in our attacks are usually close to the integer multiple of the sensor's sample rate and we have $F = n_0 \cdot F_S + {\epsilon} \ \ (|\epsilon| < 1, n_0 \in \mathbb{Z}^+ )$, where $n_0F_S$ is an integer multiple of $F_S$ that is in the resonant frequency range of the sensor.
Attackers could observe the induced actuations and estimate $|\epsilon|$. In our settings, when $|\epsilon| < 1$, the corresponding acoustic frequencies ($F$) can be considered as suitable attack frequencies.






In practice, due to sample rate drifts, $n_0F_S$ could fluctuate in a range. As a result, there could be a range of possible attack frequencies.
Since we want to use frequencies near $n_0F_S$, by tracking the range of $n_0F_S$, the range of possible attack frequencies can also be located. Attackers could try to make $|\epsilon|$ as small as possible by adjusting $F$ and estimate $n_0F_S$ from $F = n_0F_S + {\epsilon}$.

Empirically, the drift of $n_0F_S$ is usually less than 1 Hz in 1 or 2 minutes, but the accumulative drift in a long time could be larger and $n_0F_S$ could fluctuate in a frequency range with a width of around 10 Hz. We track $n_0F_S$ of the gyro in an iPhone 5 for 3 hours and find that it fluctuates in the range of 19,966 to 19,976 Hz. While it might be difficult to predict $n_0F_S$ deterministically, we notice that $n_0F_S$ tends to decrease as the target system is running, which could be caused by the increased temperature. For instance, when we just turn on a gyro-based application in an iPhone 5, $n_0F_S$ is more likely to be close to 19,975 Hz. If the application has been running for a while, $n_0F_S$ may become close to 19,970 Hz. If the application has been running for a long time such as an hour, $n_0F_S$ could be between 19,966 to 19,970 Hz.












\emph{2) Synchronizing.} Based on the profiled range of possible attack frequencies, attackers could select a frequency that is more likely to be close to $n_0F_S$ and adjust the sound frequency to `synchronize' to a suitable attack frequency to initiate the attack.

Attackers could observe changes in $|\epsilon|$ while they are adjusting $F$. Based on $F = n_0F_S + \epsilon$, if the observed $|\epsilon|$ decreases when $F$ increases, attackers could infer $F < n_0F_S$ and $\epsilon < 0$. Otherwise, they could infer $\epsilon > 0$ and $F$ should be decreased to get closer to $n_0F_S$.
In this way, attackers could adjust $F$ more effectively since they could infer the sign of $\epsilon$ and know whether the adjusted $F$ is getting closer to or further away from $n_0F_S$.






After synchronizing to a frequency $F$ with $|\epsilon|$ less than $0.5$ or $1$, attackers could start Side-Swing attacks. For Switching attacks, if attackers find a suitable $F_1$ with $ -1 < \epsilon_1 < 0$,  they could find $F_2$ by $F_2 = F_1 + 1$. Similarly, they could also acquire $F_1=F_2-1$ if they find a suitable $F_2$ with $0 < \epsilon_2 < 1$.
Usually, we make both $|\epsilon_1|$ and $|\epsilon_2|$ close to $0.5$ so that $n_0F_S$ is well bounded by $[F_1, F_2]$.



In our settings, this process involves manually tuning the acoustic frequency with an off-the-shelve function generator and observing consequent actuations of the target device. Usually, such interactions between attackers and the target system could take about 10 to 60 seconds.


\emph{3) Manipulating.} In Side-Swing attacks, attackers can increase the amplitude when the induced actuation is in the target direction and otherwise decrease the amplitude. In Switching attacks, attackers can switch the frequency of acoustic signals when the induced actuation or acceleration in the target direction begins to attenuate.


\emph{4) Adjusting (optional).} After several minutes of manipulation, $n_0F_S$ could deviate from $F$ because of sample rate drifts. Attackers could accommodate the deviation by observing changes in $\epsilon$ and adjusting $F$. For example, if attackers observe that $\epsilon < 0$ and $|\epsilon|$ increases, they could infer that $n_0F_S$ has increased and could increase $F$ to compensate for the deviation.

\vspace{-1mm}

\subsection{Experimental Setup} \label{subsection_setup}

\vspace{-0.5mm}


In our experiments, we use several types of consumer-grade tweeter speakers, including two electromagnetic (EM) speakers \cite{em_speaker2,em_speaker1} and one piezo speaker \cite{piezo_speaker}.
We measure the Sound Pressure Level (SPL) of the speakers with an NI USB-4431 sound measuring instrument and a GRAS 46AM free-field microphone that has a wide frequency range. The speaker plays single-tone sound from 1.5 kHz to 31.5 kHz with an interval of 100 Hz. We set the sample rate of the microphone to 96 kHz instead of 48 kHz to pick up ultrasonic signals correctly.

\begin{figure}
\centering
  \includegraphics[scale=0.625]{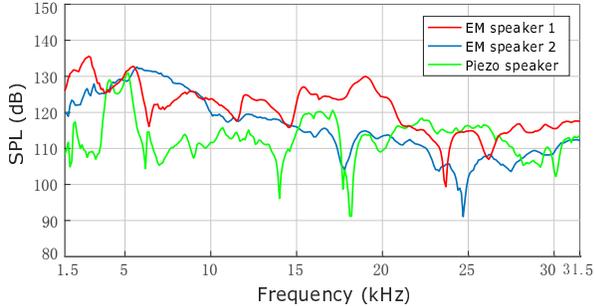}
  \vspace{-2.5mm}
  \caption{Unweighted SPL measurements of different speakers we use. The speaker is placed 10 cm from the microphone and operated near its maximum amplitude.}\label{spl_measurement}
  \vspace{-4mm}
\end{figure}






Figure \ref{spl_measurement} shows the average SPL values of the speakers, from which we can select a speaker that has the maximum SPL for each attack. The SPL of our sound source can be represented by $max(SPL_{em1},SPL_{em2},SPL_{piezo})$.
By selecting from multiple speakers, we avoid sharp performance degradations of one specific speaker in certain frequency bands and enhance the overall performance of the sound source.
The resulting improvement of SPL can be crucial in attacks on embedded sensors since the actual sound pressure grows exponentially as the sound level increases; a gain of 6.02 dB in SPL doubles the amount of sound pressure.
During attacks, we use a directivity horn, such as \cite{horn_1} and \cite{horn_2}, to improve the directivity of the sound source. The speaker is powered by a 50-Watt Lepy LP-2051 audio amplifier and the signal source is an Agilent 33220A function generator.
We conduct the experiments indoor and put acoustic foams in the environment to reduce potential sound reflections.

In Table \ref{table_close_loop} and Table \ref{table_open_loop}, we measure the maximum horizontal distance $D_{Max}$ between the sound source and the target device that an observable actuation or an inertial output with an amplitude of 0.1 rad/s can be induced under acoustic effects. Empirically, the possible attack distance with our sound source is about $\frac{D_{Max}}{4}$ for Side-Swing attacks, and $\frac{D_{Max}}{3}$ for Switching attacks to achieve adversarial control. Manufacturer information of inertial sensors is collected for statistical purposes. We find sensor information of iPhones and VR devices in online disassembling reports \cite{devices_teardown}. Android devices provide APIs to retrieve sensor information. We disassemble other devices to reveal the information written on the package of the embedded inertial sensor, but some devices do not specify the sensor model explicitly even on the sensor's package. Lastly, we record the alignments of affected and functional axes based on the orientation of the sensor when the embedded inertial sensing module is recognized. Otherwise, the alignments of axes are based on the orientation of the device.

\vspace{-1mm}

\subsection{Experiments on Closed-loop Systems}

\vspace{-0.5mm}

In a closed-loop control system, there is usually a goal state. The system continuously compares the goal state with its current state based on inertial measurements and tries to diminish the difference between them through actuations. We evaluate our attacks on different instances of four types of closed-loop systems, including self-balancing human transporters, robots, stabilizers, and anti-tremor devices. These systems present different features under acoustic effects. Nevertheless, we find that a large part of them are susceptible to implicit control.











\begin{figure}
\centering
  \includegraphics[scale=0.7]{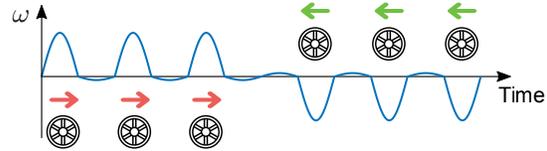}
  \vspace{-2.3mm}
  \caption{
An illustration of Side-Swing attacks on a self-balancing scooter. The system is tricked to actuate its motor based on the spoofed angular speed. The attack is demonstrated in \cite{hoverborad_video}.
}\label{sideswing_hoverboard}
  \vspace{-3.5mm}
\end{figure}

\begin{table*}
  \centering
  \caption{Results of our attack experiments on closed-loop control systems}\label{table_close_loop}
\begin{tabular}{c|c|c|c|c|c|c}
  \hline
  \multirow{2}{*}{\bf{Device}} & \multicolumn{2}{c|}{\bf{Sensor}} & \bf{Resonant} & \bf{Affected/} & \bf{Max} & \bf{Control} \\ \cline{2-3}
  & \bf{Type} & \bf{Model}${}^\dagger$ & \bf{Freq. (kHz)} & \bf{Func. Axes} & \bf{Dist. (m)} & \bf{Level}\\
  \hline \hline
   Megawheels scooter & Gyro & IS MPU-6050A & 27.1{\raise.28ex\hbox{$\scriptstyle\sim$}}27.2  & y/y & 2.9 & Implicit control\\
  \hline
Veeko 102 scooter  & Gyro & Unknown &  26.0{\raise.28ex\hbox{$\scriptstyle\sim$}}27.2   &  x/x & 2.5 & Implicit control \\
   \hline
   Segway One S1 & Gyro & Unknown & 20.0{\raise.28ex\hbox{$\scriptstyle\sim$}}20.9  & x/x & 0.8 &  Implicit control\\
  \hline
   Segway Minilite& Gyro & Unknown & 19.2{\raise.28ex\hbox{$\scriptstyle\sim$}}20.0   & x/x & 0.3 & DoS\\
  \hline
    Mitu robot & Gyro & N/A SH731 & 19.0{\raise.28ex\hbox{$\scriptstyle\sim$}}20.7   & x/x & 7.8 & Implicit Control\\
  \hline
   MiP robot & Acce & Unknown  & 5.2{\raise.28ex\hbox{$\scriptstyle\sim$}}5.4   & x/x & 1.2 & DoS \\
  \hline
  DJI Osmo stabilizer& Gyro & IS MP65  & 20.0{\raise.28ex\hbox{$\scriptstyle\sim$}}20.3  &  x,y,z/x,y,z & 1.2 & Implicit control\\
  \hline
   WenPod SP1 stabilizer& Gyro & IS MPU-6050 & 26.0{\raise.28ex\hbox{$\scriptstyle\sim$}}26.9 & z/y,z & 1.8 & Implicit control \\
  \hline
    Gyenno steady spoon & Gyro & Unknown  & Not found   & Unknown & N/A & Not affected\\
    \hline
    Liftware level handle & Acce & IS MPU-6050   & 5.1   & x/x & 0.1 & DoS\\
    \hline
\end{tabular}
    \begin{tablenotes}
      \small
      \item  $\dagger$ IS: InvenSense, N/A: Unknown manufacturer.
    \end{tablenotes}
\vspace{-3.5mm}
\end{table*}

\smallsection{(1) Human transporters}
The goal state of self-balancing human transporters is a vertical position of the system with a tilt angle of 0$^\circ$. Inertial sensors are used to detect tilts of the transporter. Based on the direction and amount of the tilt, the control system applies accelerations to motors to correct the position of the system.

We evaluate acoustic attacks on four instances of self-balancing transporters: a Megawheels TW01 scooter, a Veeko 102 scooter, a Segway one S1 unicycle, and a Segway Minilite scooter. We find that, by spoofing the angular speed measured by gyros, the moving direction and speed of the motor could be controlled, as illustrated in Figure \ref{sideswing_hoverboard}.


\paragraph{Results} The Megawheels scooter and the Veeko scooter are vulnerable to adversarial control over the moving direction and speed of the motor through ultrasonic signals.
While the Segway One S1 unicycle can be manipulated by Switching attacks, the range of induced actuations is very small. The unicycle only tilts slightly to the target direction. The Segway Minilite scooter tends to lose control under acoustic effects. Our Side-Swing attacks and Switching attacks on smart human transporters are demonstrated in \cite{hoverborad_video} and \cite{hoverboradswitching_video}\footnotemark. The transporter is in a relatively static experimental setting, and we lift the wheels of the transporter up from the ground during the experiments.

\footnotetext{Precautions were used to ensure the safety of researchers.}

\smallsection{(2) Robots} Self-balancing robots work similarly to self-balancing human transporters but without a rider. We test two self-balancing robots equipped with MEMS gyros and accelerometers: a Mitu robot and a MiP robot.

\paragraph{Results} We find that the gyro of Mitu robot is susceptible to adversarial control. The robot would speed up to the same direction as the spoofed rotations under Side-Swing attacks, as demonstrated in \cite{mitu_video}.
While the gyro of MiP robot is not affected by acoustic attacks, its accelerometer is vulnerable to DoS attack, which makes it suddenly stop working and fall to the ground.







\smallsection{(3) Stabilizers}
MEMS inertial sensors are widely used in aiming and stabilizing systems.
The goal of such systems is to maintain a device or platform in a certain orientation despite external forces or movements.
Therefore, when movements are detected by inertial sensors, the system would actuate in opposite directions to cancel the effect of external movements.

\begin{figure}
\centering
  \vspace{-2mm}
  \includegraphics[scale=0.95]{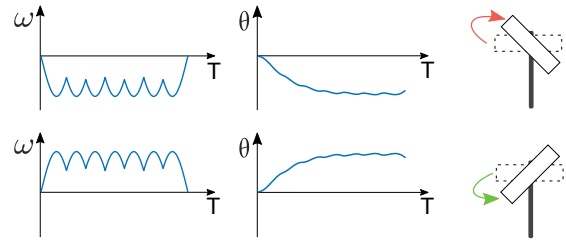}
  \vspace{-2mm}
  \caption{
An illustration of Switching attacks on a stabilizer. The stabilizer tries to correct the fabricated heading angle in Y-axis of the device by rotating to the opposite direction. The attack is demonstrated in \cite{stabilizer_video}.
}\label{switching_stabilizer}
  \vspace{-4mm}
\end{figure}

We evaluate our attacks on two camera stabilizers: a DJI Osmo stabilizer and a Wenpod SP1 stabilizer.
Our results show that by spoofing the gyro and manipulating the derived heading angle, the pointing direction of a stabilizer could be controlled. However, fabricated heading angles in X and Y axes will be gradually calibrated by the system based on gravity information. As illustrated in Figure \ref{switching_stabilizer}, we can use Switching attacks to induce a maximum heading degree in the stabilizer. As the induced heading angle increases, the calibration effect also becomes stronger until the maximum heading angle is reached.





\paragraph{Results} Both instances of stabilizers are vulnerable to adversarial control through ultrasonic signals. The Osmo stabilizer is mainly affected in X-axis while the Wenpod stabilizer can only be manipulated in Y-axis of the device (which is the Z-axis based on the orientation of the embedded inertial sensor). Our Side-Swing attacks and Switching attacks on stabilizers are demonstrated in \cite{stabilizer_video2_swing} and \cite{stabilizer_video}.








\smallsection{(4) Anti-tremor Devices} Inertial sensors can be used by anti-tremor gadgets in health-care applications, such as gyroscopic tablewares and gloves \cite{mit_glove} that mitigate hand tremors and assist users to perform daily tasks. We evaluate acoustic attacks on a Liftware level handle and a Gyenno gyroscopic spoon.


\paragraph{Results} The Liftware handle is vulnerable to DoS attacks on its accelerometer. The handle under attacks would abnormally actuate its motor to one direction and become unusable. 
The Gyenno gyroscopic spoon is not affected by acoustic signals.




\subsection{Experiments on Open-loop Systems}


Different from closed-loop systems that have a goal state, open-loop control systems simply take inertial measurements as inputs and actuate accordingly.
We evaluate our attacks on various types of devices that use real-time inertial data for open-loop control.
These devices use various MEMS inertial sensors from different vendors. Nevertheless, we find that most of them could be susceptible to implicit control.








\smallsection{(1) 3D mouses}
Inertial sensors can be used in input devices for remote control.
3D mouses use gyros to detect a user's hand movements and move the cursor accordingly. We evaluate our spoofing attacks on an IOGear 3D mouse and a Ybee 3D mouse.

\vspace{-0.5mm}
\paragraph{Results}
Both instances of 3D mouse are vulnerable to adversarial control through ultrasonic signals.
By spoofing the gyroscope, attackers could point the cursor of the 3D mouse in a remote system to different targets.
We demonstrate Side-Swing attacks and Switching attacks on 3D mouses in \cite{3dmouse_swing} and \cite{3dmouse_switching}.


\smallsection{(2) Gyroscopic screwdrivers} The gyroscopic screwdriver is an industrial application that controls a mechanical system based on inertial measurements. The moving direction and speed of the motor in the screwdriver is decided by the heading angle derived from gyroscope data.





\begin{figure}
\centering
  \includegraphics[scale=0.82]{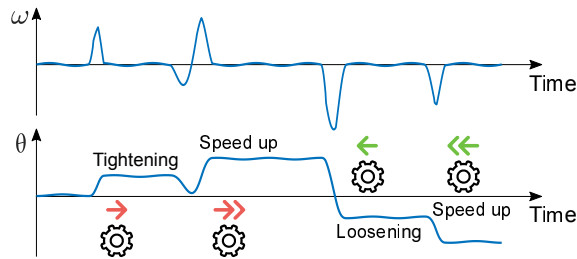}
  \vspace{-1.6mm}
  \caption{
An illustration of Conservative Side-Swing attacks on a screwdriver. Both the moving direction and speed of the motor can be manipulated by spoofing the gyroscope. The attack is demonstrated in \cite{screwdriver_video}.
}\label{swing_screwdriver}
  \vspace{-3.9mm}
\end{figure}

In gyroscopic screwdrivers, there is usually no mechanism to calibrate the heading angle. Therefore, the induced heading angle will not be eliminated even when the attack ceases.
Based on this feature, we adjust our attack method to \emph{Conservative Side-Swing Attacks}. The basic idea is that attackers emit acoustic signals only when changing the direction or speed of the motor. Once the motor is tricked to move with a desired speed in the target direction, attackers can turn off acoustic signals to keep the heading angle in the system, as illustrated in Figure \ref{swing_screwdriver}.
We evaluate our attacks on an E-design ES120 screwdriver, a B\&D gyroscopic screwdriver, and a Dewalt gyroscopic screwdriver.





\vspace{-0.5mm}
\paragraph{Results} By spoofing the gyro and manipulate the derived heading angle, both the moving direction and speed of the motor in the ES120 screwdriver can be controlled. The B\&D screwdriver can be manipulated only after we remove its external panel and the Dewalt screwdriver is not affected by acoustic signals.


\smallsection{(3) VR/AR devices}
Inertial sensors are used by Virtual/Augmented Reality (VR/AR) headsets and kinetic controllers to track the user's movements and control visual information in an image system. The user's view in VR systems or the position of augmented information displayed in AR systems is often determined by heading angles of the headset. In addition, the movements detected by the kinetic controller will directly be used to control an object in the image system. We evaluate our attacks on an Oculus Rift VR headset, an Oculus Touch controller, and a Microsoft Hololens AR headset.

\vspace{-0.7mm}
\paragraph{Results}
By spoofing the gyros with ultrasonic signals, the user's view in Oculus Rift headset and the orientation of an object controlled by Oculus Touch can both be manipulated in X-axis. The Hololens headset can only be affected very slightly by our sound source. Our Switching attacks on VR devices are demonstrated in \cite{touch_switching} and \cite{rift_switching}.
Recent researches have shown that buggy or maliciously exploited visual information in an immersive environment might startle or mislead a user and cause unexpected consequences \cite{lebeck2017securing,lebeck2018towards}.
Furthermore, a few prototype products use AR applications to assist critical real-world tasks \cite{ar_army,ar_driving}, and plenty of studies utilize inertial measurements to remotely control mechanical systems such as a robotic arm \cite{bhuyan2014gyro}. Our experimental results might help designers of these rapidly emerging applications to be aware of potential threats that might be caused by spoofing inertial sensors.

\begin{table*}
  \centering
  \begin{threeparttable}
  \caption{Results of our attack experiments on open-loop control systems}\label{table_open_loop}
\begin{tabular}{c|c|c|c|c|c|c}
  \hline
  \multirow{2}{*}{\bf{Device}} & \multicolumn{2}{c|}{\bf{Sensor}} & \bf{Resonant} & \bf{Affected/} & \bf{Max} & \bf{Control} \\ \cline{2-3}
  & \bf{Type} & \bf{Model}$^\dagger$ & \bf{Freq. (kHz)} & \bf{Func. Axes} & \bf{Dist. (m)} & \bf{Level}\\
  \hline \hline
   IOGear 3D mouse & Gyro   &  IS M681  & 26.6{\raise.28ex\hbox{$\scriptstyle\sim$}}27.6 & x,z/x,z & 2.5 & Implicit control\\
  \hline
   Ybee 3D mouse & Gyro   &  Unknown   & 27.1{\raise.28ex\hbox{$\scriptstyle\sim$}}27.3 & x/x,z & 1.1 & Implicit control\\
  \hline
ES120 screwdriver & Gyro & ST L3G4200D &  19.8{\raise.28ex\hbox{$\scriptstyle\sim$}}20.0 & y/y & 2.6 & Implicit control \\
  \hline
B\&D screwdriver  & Gyro & IS ISZ650    & 30.3{\raise.28ex\hbox{$\scriptstyle\sim$}}30.6  & z/z  &  0 & Limited control \\
    \hline
Dewalt screwdriver  & Gyro & Unknown   & Not found  & none/y  &  N/A & Not affected \\
    \hline
Oculus Rift    & Gyro  & BS BMI055   & 24.3{\raise.28ex\hbox{$\scriptstyle\sim$}}25.6   & x/x,y,z & 2.4 & Implicit control \\
  \hline
Oculus Touch
 & Gyro & IS MP651  &  27.1{\raise.28ex\hbox{$\scriptstyle\sim$}}27.4  & x/x,y,z & 1.6 & Implicit control \\
  \hline
Microsoft Hololens    & Gyro  & Unknown & 27.0{\raise.28ex\hbox{$\scriptstyle\sim$}}27.4   & x/x,y,z & 0 & Limited control \\
  \hline
  iPhone 5 & Gyro &  ST L3G4200D   & 19.9{\raise.28ex\hbox{$\scriptstyle\sim$}}20.1
  & x,y,z/x,y,z & 5.8 & Implicit control\\
  \hline
  iPhone 5S & Gyro &  ST B329  & 19.4{\raise.28ex\hbox{$\scriptstyle\sim$}}19.6 & x,y,z/x,y,z & 5.6 & Implicit control\\
  \hline
  iPhone 6S & Gyro & IS MP67B    & 27.2{\raise.28ex\hbox{$\scriptstyle\sim$}}27.6  & x,y,z/x,y,z &  0.8 & Implicit control \\
  \hline
  iPhone 7 & Gyro &  IS 773C &   27.1{\raise.28ex\hbox{$\scriptstyle\sim$}}27.6 & x,y,z/x,y,z & 2.0 & Implicit control\\
  \hline
  Huawei Honor V8 & Gyro & ST LSM6DS3    & 20.2{\raise.28ex\hbox{$\scriptstyle\sim$}}20.4
 & x,y,z/x,y,z & 7.7  & Implicit control\\
  \hline
  Google Pixel & Gyro & BS BMI160    & 23.1{\raise.28ex\hbox{$\scriptstyle\sim$}}23.3  &  x,y,z/x,y,z & 0.4 & Implicit control\\
  \hline
Pro32 soldering iron   & Acce & NX MMA8652FC    &  6.2{\raise.28ex\hbox{$\scriptstyle\sim$}}6.5  &   Unknown   & 1.1 & DoS  \\
  \hline
\end{tabular}
    \begin{tablenotes}
      \small
      \item  $\dagger$ IS: InvenSense, ST:STMicroelectronics, BS: Bosch, NX: NXP Semiconductors.
    \end{tablenotes}
  \end{threeparttable}
\vspace{-3mm}
\end{table*}

\vspace{-0.2mm}
\smallsection{ (4) Smartphones} Smartphones have become a platform that provides sensor data and computation resource for large amounts of applications. Inertial sensor data of smartphones is often used in mobile VR/AR applications and navigation systems. We evaluate our attacks on six smartphones in different models. Both iOS and Android devices are tested.


%

\vspace{-0.5mm}
\paragraph{Results} The smartphones we test have different gyroscopes, which have different resonant frequency ranges. While their sensitivity to resonant sound differs, we find that all of them are vulnerable to adversarial control.
Our Side-Swing attacks and Switching attacks on mobile VR applications are demonstrated in \cite{phone_swing} and \cite{phone_switching}. In the demos, we manipulate the VR user's view and aim several targets by spoofing the gyroscopic sensor.

\vspace{-0.2mm}
\smallsection{(5) Motion-aware devices} Using inertial sensors to detect motions is a popular wake-up mechanism in smart devices. This mechanism can also be used to control critical functions of an embedded system. The Pro32 soldering iron uses an accelerometer to detect movements. If there is no movement for a long time, the system will cool down the iron tip and go into the sleep mode. This protects the iron from overheating and reduces the risk of accidental injuries or fire.
However, we find that this mechanism could be compromised by resonant acoustic interferences.
Our experiments show that attackers can wake the Pro32 soldering iron up from the sleep mode through DoS attacks on the accelerometer, and make the iron tip heat up to a high working temperature repetitively. The attack is demonstrated in \cite{solderingiron_video}.











\vspace{-2mm}
\section{Automatic Attack} \label{section_automatic}
\vspace{-1mm}

In this section, we present a novel automatic attack method and implement a proof-of-concept spoofing attack on a mobile navigation system.
We find that in both iOS and Android smartphones, inertial sensor data can be accessed through a script in a web page or an application without any permission. In our scope, a key question is: \emph{Can an attack program facilitate spoofing attacks on inertial sensors by leveraging the real-time sensor data?} To answer this question, we investigate automatic methods to implement Switching attacks.

\vspace{-0.5mm}

\paragraph{Automatic Method} In automatic attacks, the attack program modulates acoustic signals automatically based on parameters set by the attacker. These parameters include initial sound frequencies, threshold, target direction, etc. The attacker can set the initial sound frequencies $F_1$ and $F_2$ based on the real-time feedback of the sensor. The threshold is used by the attack program to decide when to switch the sound frequency. During attacks, the attacker can send commands to the program to change the target direction, to stop or restart the attack.












The attack program monitors the output of the sensor and switches the frequency of acoustic signals between $F_1$ and $F_2$ when the induced signal drops to the opposite direction and falls below a threshold.
However, we find that this setting only allows the program to attack automatically for one or two minutes. After two minutes, the integer multiple of the sensor's sample rate might fall outside ($F_1$, $F_2$) because of drifts in $F_S$ and the condition of phase pacing ($\epsilon_1 \cdot \epsilon_2 < 0$) would no longer hold. As a result, the attacker would need to manually adjust the sound frequencies every one or two minutes.


A method to address this issue is to \emph{actively adapt to the drifts in the sample rate}.
Due to drifts in $F_S$, the value of $n_0F_S$ may become $n_0\hat{F_S}$. If $n_0\hat{F_S}$ falls outside $(F_1, F_2)$, the condition of phase pacing will no longer be satisfied.
 Therefore, the goal of adaptation is to actively adjust the sound frequencies to $\hat{F_1}$ and $\hat{F_2}$ so that $n_0\hat{F_S}$ is at the midpoint of $(\hat{F_1}, \hat{F_2})$.
Assuming $\epsilon_1 < 0, \epsilon_2 > 0$, we have,

\vspace{-2mm}
\begin{equation}
\begin{array}{l}

F_1 - \epsilon_1 = n_0\hat{F_S} = F_2 - \epsilon_2

\end{array}
\vspace{-1mm}
\end{equation}

After adaptation, we would have,

\vspace{-2mm}
\begin{equation}
\begin{array}{l}

\hat{F_1} + \frac{\epsilon_2 - \epsilon_1}{2} = n_0\hat{F_S} = \hat{F_2} - \frac{\epsilon_2 - \epsilon_1}{2}

\end{array}
\vspace{-1mm}
\end{equation}


Therefore,

\vspace{-4mm}
\begin{equation}
\begin{array}{l}

\Delta F = \hat{F_1} - F_1 = \hat{F_2} - F_2 = -\frac{\epsilon_1 + \epsilon_2}{2(\epsilon_2 - \epsilon_1)}(\epsilon_2 - \epsilon_1)

\end{array}
\vspace{-1mm}
\end{equation}


Since $\epsilon_2 - \epsilon_1 = F_2 - F_1$, we have,

\vspace{-2mm}
\begin{equation}
\begin{array}{l}

\Delta F = \frac{r - 1}{2(r + 1)}(F_2 - F_1)

\end{array}
\vspace{-1mm}
\end{equation}


where $r=\frac{|\epsilon_1|}{|\epsilon_2|}=\frac{-\epsilon_1}{\epsilon_2}$, and can be derived from



\vspace{-2mm}
\begin{equation}
\begin{array}{l}

r = \frac{T_2}{T_1} \approx \frac{T'_2}{T'_1}

\end{array}
\vspace{-1mm}
\end{equation}

$T_1$ and $T_2$ are periods of the induced signals. The ratio $\frac{T_2}{T_1}$ can be estimated by $\frac{T'_2}{T'_1}$, where
$T'_1$ and $T'_2$ correspond to the time intervals between adjacent frequency switching operations. During attacks, $T'_1$ and $T'_2$ can be recorded by the program. The program computes $\Delta F$ and adapts the frequencies after every two times of frequency switching.








\paragraph{Evaluation} We evaluate our attacks on a Huawei Honor V8 smartphone and demonstrate the attack effects with a mobile navigation system (Google Maps).
In mobile navigation systems, inertial sensors are often used to aid the GPS system to provide a more timely and accurate positioning service.
The gyroscope is often used to determine the orientation of the system.



\begin{figure}
\centering
  \includegraphics[scale=0.69]{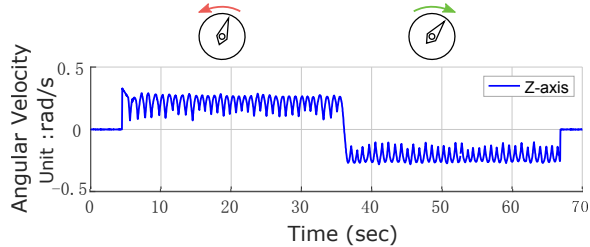}
  \vspace{-3mm}
  \caption{
  Controlling the orientation of a mobile navigation system with automatic Switching attacks on the gyroscopic sensor. The attack is demonstrated in \cite{automatic_video}.
}\label{autoswitching_map}
  \vspace{-5mm}
\end{figure}

We implement the automatic attack method in an Android application. The application utilizes the smartphone's built-in speaker to generate ultrasonic signals and surreptitiously manipulate the gyroscope data while running in the background. As shown in Figure \ref{autoswitching_map}, we first induce positive outputs in the Z-axis of gyro and the navigation system is tricked to rotate its orientation counter-clockwisely. The accumulated heading angle is 6.85 rads in 32 seconds.
After we change the target direction, the navigation system is deceived by negative outputs and rotates the orientation clockwisely. The accumulated heading angle is -6.82 rads in about 31 seconds.







Our results show that, with real-time sensor data, spoofing attacks on inertial sensors could manipulate the orientation of a navigation system.  When the displayed orientation of a navigation system is manipulated, users or systems guided by the navigation information could be led to a wrong path. Additionally, for areas not well covered by GPS or situations when the GPS signal is jammed or spoofed \cite{nighswander2012gps,psiaki2014gnss}, errors in the orientation information will not be effectively calibrated and could cause more troubles to the positioning service.







Several recent approaches have been proposed to control the access to inertial sensors in smartphones, but with a focus on privacy issues \cite{petracca2017aware,sikder20176thsense}.
Our automatic attack also demonstrates that unprotected inertial sensor data could be leveraged to manipulate the sensor output.
Our results confirm that protection mechanisms over inertial sensor data are necessary.
Devices should control the access to the sensor data. In addition, when a remote autonomous agent transmits real-time inertial sensor data for navigation purposes, the data should be encrypted.

\vspace{-0.5mm}

\section{Discussion}
\vspace{-0.5mm}


\subsection{Countermeasures}


It is important to protect control systems from sensor spoofing attacks, however, feasible countermeasures to be deployed in embedded systems should not cause too much expenses in cost and size or compromises in designs.
Therefore, the countermeasures we discuss mainly focus on two aspects: (1) Damping and isolation. These approaches mitigate acoustic or vibrational noises physically. (2) Filtering and sampling. These approaches eliminate or mitigate malicious signals in the signal conditioning circuits.

\paragraph{Damping and Isolation} 
Early mitigation approaches against acoustic interferences include using isolating boxes and acoustic foams to surround the sensor \cite{castro2007influence}. The simple strategy could achieve substantial protection from acoustic noises, but issues in size and design concerning an embedded environment were not addressed.

To protect MEMS inertial sensors without compromising their advantages in size, weight, power, and cost (SWaP-C \cite{kranz2017environmentally}), recent studies have been dedicated to using micro-level techniques for acoustic isolation.
Dean et al. proposed the use of microfibrous metallic cloth as an acoustic damping material to protect MEMS gyroscopes \cite{dean2011microfibrous}.
Soobramaney et al. evaluated the mitigation effects of microfibrous cloth on noise signals induced in MEMS gyros under acoustic interferences \cite{soobramaney2015mitigation}.
They tested 7 MEMS gyros and showed that,
by surrounding the sensor with 12 mm of the media, 65\% reduction in the amplitude of noise signals can be easily obtained and up to 90\% reduction could be achieved \cite{soobramaney2015mitigation}.
Additionally, Yunker et al. suggested to use MEMS fabricated acoustic metamaterial to mitigate acoustic signals at frequencies close to the resonant frequency of the MEMS gyroscope \cite{yunker2013sound}.
Furthermore, Kranz et al. showed that a MEMS-fabricated micro-isolator can be applied within the sensor packaging but their work mainly focused on isolating mechanical vibrations \cite{kranz2017environmentally}.

\paragraph{Filtering}
As suggested in \cite{trippel2017walnut}, a low-pass filter (LPF) should be used to eliminate the out-of-band analog signals.
According to the datasheets \cite{l3gd20_datasheet,lsm330_datasheet}, we find that many inertial sensors have an analog LPF in their circuits, but are still vulnerable to acoustic attacks, which could be due to a cut-off frequency that is set too high.
We also find that most programmable inertial sensors use a digital LPF for bandwidth control \cite{l3g4200_datasheet,mpu_6500_datasheet}.
However, filters in digital circuits will not alleviate the problem because out-of-band analog signals have already been aliased to in-band signals after sampling.



\paragraph{Sampling} Trippel et al. proposed randomized sampling and 180$^\circ$ out-of-phase sampling methods for inertial sensors with analog outputs and software controlled ADCs \cite{trippel2017walnut}. These approaches were designed to eliminate an attacker's ability to achieve a DC signal alias and limit potential adversarial control. However, adding a randomized delay to each sampling period or computing the average of two samples at a 180$^\circ$ phase delay could degrade the accuracy of inertial measurements. Small errors in the measurements could accumulate in a long time and might affect the performance of the system.


We think an alternative sampling method to mitigate potential adversarial control without degrading the performance is to use a \emph{dynamic sample rate}.
Recall in (\ref{eq4_3_pre_model}) and (\ref{eq4_4_basic_model_equation}), the frequency $\epsilon$ of the induced digital signal depends on both $F$ and $F_S$. With a dynamic $F_S$, attackers may not be able to induce a digital signal with a predictable frequency pattern.
In this case, the ability of attackers will be limited and it could be difficult for attackers to accumulate a large heading angle in a target direction.
This might be a general mitigation method for ADCs subject to out-of-band signal injections.


Additionally, redundancy-based approaches could enhance the resilience of the system. For example, multiple sensors could still provide trustworthy information when one of them is under attack. It might still be possible to attack or interfere several sensors simultaneously to affect the functioning of the system, but such attacks could be more difficult to implement.


In summary, acoustic attacks on inertial sensors are enabled by two weaknesses in the analog domain: (1) Susceptibility of the micro inertial sensing structure to resonant sound.
(2) Incapability of signal conditioning circuits to handle out-of-band analog signals properly.
Employing both acoustic damping and filtering approaches in the designs of future sensors and systems can address these weaknesses. Additionally, acoustic damping can also be used to mitigate the susceptibility of currently deployed sensors and systems to acoustic attacks.

\subsection{Sound Source}

Applications of sonic weapons \cite{altmann2001acoustic}, ultrasonic transducers \cite{gallego1978ultrasonic}, and long-range acoustic devices \cite{lrad,hypersonic} have already shown the capability of specialized devices to generate more powerful sound with a further transmitting distance than common audio devices.
In addition, we find several consumer-grade techniques that could be used to optimize a sound source.


The most direct acoustic amplification method is to use speakers and amplifiers with better performance and output capabilities.
Besides, the sound played by common audio speakers usually diffuses into the air with little directivity, leading to losses of acoustic energy. With directivity horns \cite{horn_1,horn_2}, the sound waves can be focused into a certain emitting area and transmit through a longer distance.
Another important approach is to use multiple speakers to form a specialized speaker array. With appropriate arrangement of speakers and directivity horns to focus the sound waves, the sound strength, transmitting distance, and emitting area of the sound source could be customized and improved.
Moreover, ultrasonic transducers \cite{wygant200950,wang2006micromachined} could have small sizes, variable resonant frequencies, and high efficiency. It might be possible to build a more powerful and efficient sound source by selecting and using a large number of transducers.

With multiple speakers or transducers, the performance of a sound source could be improved. If the sound waves are in phase, the add-up SPL of $n$ coherent sources could be \cite{add_spl},

\vspace{-4mm}
\begin{equation}
\begin{array}{l}

L_{\Sigma} = 20log_{10}(10^{\frac{L_{p1}}{20}} + 10^{\frac{L_{p2}}{20}} + ... + 10^{\frac{L_{pn}}{20}})

\end{array}
\vspace{-1.2mm}
\end{equation}

Assuming each coherent source is identical, we have

\vspace{-2mm}
\begin{equation}
\begin{array}{l}

L_{\Sigma} = 20log_{10}(n) + L_{p1}

\end{array}
\vspace{-0.8mm}
\end{equation}



Theoretically, with 8 identical sources, the level increase could be $L_{\Sigma}-L_{p1}\approx 18.0$ dB. In practice, the performance could also depend on arrangements of multiple sources, designs of the enclosure and horns, and differences in phases need to be considered and accommodated.
The distance attenuation of SPL can be quantified by \cite{distance_spl}:
$L'_{p}=L_{p}+20log_{10}(\frac{D}{D'})$, where $D$ and $D'$ are distances. Therefore, a level increase of $18.0$ dB could increase the possible attack distance by a factor of 8.












\subsection{Limitations}

\paragraph{Moving targets} Depending on the speed and range of movements, it could be difficult for attackers to follow and aim a moving target while manually tuning acoustic signals. It could be helpful to predict the movements and align the sound beam with the trajectory of the target.
In certain circumstances, it might be possible to attach a sound source to the victim device or exploit a sound source in close proximity to the device. Additionally, it might be possible to carry the sound source with a vehicle or drone that follows the target.

Ideally, an automatic tracking and aiming system might be implemented to aim the target. It might use cameras or radar sensors to track the position of a target and use a programmable 3-way pan/tilt platform to aim.





\paragraph{Timing} In our experimental settings, attackers observe actuations of a target and manually tune acoustic signals with off-the-shelve devices.
In certain circumstances, however, such settings could be slow and ineffective; it might be difficult for attackers to analyze the observed movements and modulate signals timely and correctly.

To reduce potential delays caused by hand tuning and observing, it might be possible to use more customized devices, tools, and programs.
As we have investigated in Section \ref{section_automatic}, a program could help attackers to modulate acoustic signals more timely and accurately. Moreover, it might also be possible to use systems with cameras or radar sensors to help attackers observe and analyze the behavior of a target more automatically.








In addition, the pattern of a closed-loop system could be more complex than the simple signal mapping model in Section \ref{subsection_overview}. For example, when a user is riding the self-balancing scooter, user involvements (including unintentional involvements) could counter or disrupt attack effects. Attackers might need a more specific model to analyze and predict the movement patterns.

\subsection{Generalization} \label{sec_generalization}


Acoustic attacks on inertial sensors exploit resonance and inject analog signals with very high frequencies. To explore the generalizability of the out-of-band signal injection model and attack methods, we investigate whether the oscillating digitized signal could be manipulated when analog signals are sent at relatively low frequencies through a more common sensing channel.

\begin{figure}
\centering
  \includegraphics[scale=0.70]{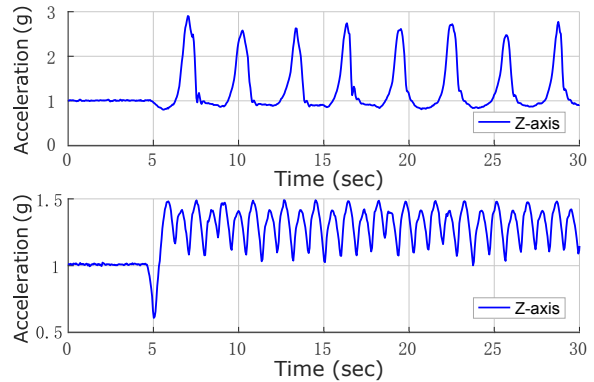}
  \vspace{-3mm}
  \caption{
  The output of the accelerometer (Z-axis) in a Google Pixel smartphone. We implement Side-Swing (top) and Switching attacks (bottom) with low-frequency vibration signals to manipulate the sensor output. The phone is placed with the Z-axis pointing upward, and the default output in Z-axis is 1 g if the device is at rest.
}\label{acc_swing}
  \vspace{-3mm}
\end{figure}

We use a vibrating platform to generate mechanical vibration signals and implement Side-Swing and Switching attacks on the accelerometer of a smartphone, as shown in Figure \ref{acc_swing}.
We place the Google Pixel smartphone on the platform. In Side-Swing attacks, we generate sinusoidal vibration signals at 19.6 Hz. While the phone remains on the platform, the collected accelerometer data shows that the phone is launched to the sky and has accumulated a speed of 73.9 m/s in about 25 seconds. In Switching attacks, we switch the frequency of the sinusoidal vibration signal between 19.4 Hz and 20.4 Hz. While the phone is still placed on the platform, the accelerometer data shows that it has accumulated an upward speed of 74.5 m/s in about 25 seconds.

We try to find the approximate sample rate of the embedded accelerometer by inducing an aliased DC-like signal. We increase the vibration frequency with an interval of 0.1 Hz and observe the induced output. The first DC-like signal is induced at $F=19.9$ Hz, the second at 39.8 Hz, and the third at 59.7 Hz. Based on $F=nF_S+\epsilon_0 \ \ (\epsilon_0 \approx 0)$, we infer that the sample rate of the ADC is approximately 19.9 Hz.




Our experimental results show that, when analog signals are sent at relatively low frequencies, such as frequencies close to $F_S$,
the oscillating digitized signal could still be manipulated. Moreover, instead of exploiting resonance, malicious signals could be injected and manipulated through the sensing channel as well.

As we have discussed, sensors without a correctly configured analog LPF could be vulnerable to out-of-band signal injections.
Furthermore, some digital sensors could have a configurable sample rate and use a programmable digital LPF for bandwidth control.
For example, the ADC sample rate of the MPU-6500 gyroscope is programmable from 8,000 samples per second, down to 3.9 samples per second \cite{mpu_6500_datasheet}.
In this case, assuming the cut-off frequency of the analog LPF is 4 kHz, which is the half of the maximum sample rate, if applications set $F_S$ to 4 kHz or lower, out-of-band signals with relatively low frequencies (such as frequencies close to $F_S$) would not be eliminated by the analog LPF and could be exploited to manipulate the digitized signal.

\vspace{-2mm}

\section{Related Work} \label{section_rw}
\vspace{-1mm}

Since measurements of embedded sensors are often trusted by control systems to make critical decisions, the security of analog sensors has become an increasingly important concern. This section discusses security of inertial sensors and attacks against analog sensors.


\paragraph{Attacks on Inertial Sensors} MEMS inertial sensors have drawn the attention of recent security researches because of their criticality in control systems. Son et al. \cite{son2015rocking} proposed a DoS attack against MEMS gyroscopes and showed that a drone could be caused to crash by intentional resonant sound. Additionally, Wang et al. developed a sonic gun and showed that a range of smart devices could lose control under acoustic attacks on inertial sensors \cite{wang2017gun}.
Furthermore, Trippel et al. \cite{trippel2017walnut} proposed output biasing attacks and output control attacks to compromise the integrity of MEMS accelerometers. However, output biasing attacks were only implemented on exposed sensors with an insufficiently realistic attack setting; while the output control attack method only works on sensors with an insecure amplifier and the generalizability could be limited in two aspects: (1) To trigger signal clipping in the amplifier, the amplitude of the induced analog signal needs to exceed the operating range of the amplifier. (2) The direction of induced outputs is determined by the asymmetricity of signal clipping that occurs in the saturated amplifier and cannot be controlled.
Different from prior works, this work shows that an oscillating digitized signal, which is often regarded as noises, could be manipulated to deliver adversarial control, and demonstrates implicit control over different kinds of real systems through non-invasive attacks against embedded inertial sensors.

\paragraph{Eavesdropping through Inertial Sensors} Inertial sensors have become ubiquitous in mobile devices. 
It is known that access to inertial sensors in both iOS and Android devices does not require permissions from the operating system~\cite{cai2012practicality,michalevsky2014gyrophone}.
Therefore, attackers could surreptitiously read inertial sensor data through either a web script or a malicious application.
The inertial sensing data in smartphones could be used to recover keystroke information \cite{cai2012practicality,aviv2012practicality, miluzzo2012tapprints}.
Furthermore, the works of \cite{michalevsky2014gyrophone} and \cite{anand2018speechless} showed that it might be possible to utilize inertial sensors in a smartphone to eavesdrop speech signals in certain scenarios.
Additionally, a user's keystroke information could be recovered by exploiting inertial sensors in smart watches \cite{liu2015good, wang2016friend, wang2015mole}. More recent studies showed that inertial sensors in mobile devices could be exploited to establish a covert channel due to their sensitivity to vibrations \cite{farshteindiker2016phone,block2017autonomic}.
All these works focused on utilizing inertial sensing data for eavesdropping or data exfiltration purposes. To our knowledge, the automatic attack we demonstrate is the first method that leverages inertial sensor data to manipulate the sensor output with a malicious program.





\paragraph{Analog Sensor Spoofing Attacks} Foo Kune et al. showed that bogus signals could be injected into analog circuits of a sensor through electromagnetic interference to trigger or inhibit critical functions of cardiac implantable electrical devices \cite{kune2013ghost}. Park et al. studied a saturation attack against infrared drop sensors to manipulate the dosage delivered by medical infusion pumps \cite{park2016ain}.
In automotive embedded systems, Shoukry et al. presented non-invasive spoofing attacks on magnetic wheel speed sensors in anti-lock braking systems \cite{shoukry2013non}. Yan et al. investigated contactless attacks against environment perception sensors in autonomous vehicles \cite{yan2016can}. Recently, Shin et al. studied spoofing attacks on Lidar sensors in automotive systems to manipulate the distance of objects detected by the system \cite{shin2017illusion}.
In addition, Davidson et al. investigated a sensor input spoofing attack against optical flow sensing of unmanned aerial vehicles \cite{davidson2016controlling}.
Finally, Zhang et al. presented an inaudible attack on voice controllable systems that injects commands into a microphone through ultrasonic carriers \cite{zhang2017dolphinattack}.

\vspace{-1mm}

\section{Conclusion}
\vspace{-0.5mm}


Embedded sensors in a control loop play important roles in the correct functioning of control systems. A wide range of control systems depend on the timely feedback of MEMS inertial sensors to make critical decisions.
In this work, we devised two sets of novel attacks against embedded inertial sensors to deceive the system. Our attack evaluations on 25 devices showed that it is possible to deliver implicit control to different kinds of systems by non-invasive attacks.


We characterized the out-of-band signal injection model and methods to manipulate an oscillating digitized signal, which was often considered as noises, to deliver adversarial control. To explore the generalizability of our methods, we showed that the oscillating digitized signal could also be manipulated by sending analog signals at relatively low frequencies through the sensing channel.

\section*{Acknowledgment} \vspace{-2pt}

The authors would like to thank the anonymous reviewers and our shepherd Yongdae Kim for their numerous, insightful comments that greatly helped improve the presentation of this paper. This work is supported in part by ONR N000141712012 and US NSF under grants CNS-1812553, CNS-1834215, and CNS-1505799.